\documentstyle[aas2pp4]{article}

\received{29 June 2000}
\accepted{12 September 2000}
\journalid{337}{02 December 1998}
\articleid{11}{14}

\slugcomment{\bf{To appear on The Astrophysical Journal}}

\lefthead{M\'endez and Ruiz}
\righthead{The luminosity function of white-dwarfs.}

\begin{document}

\title{The Luminosity Function of Magnitude and Proper-Motion Selected
Samples.\\ The case of White-Dwarfs.}

\author{Ren\'e A. M\'endez}
\affil{Cerro Tololo Inter-American Observatory, Casilla 603, La
Serena, Chile. \\ E-mail: rmendez@noao.edu}

\author{Mar\'{\i}a Teresa Ruiz}

\affil{Departamento de Astronom\'{\i}a, Universidad de Chile, Casilla
36-D, Santiago, Chile. \\ E-mail: mtruiz@das.uchile.cl}

\author{\bf{To appear on \sc{The Astrophysical Journal}}}

\begin{abstract}

The luminosity function of white dwarfs is a powerful tool for studies
of the evolution and formation of the Milky Way. The (theoretical)
white dwarf cooling sequence provides a useful indicator of the
evolutionary time scales involved in the chronometry and star
formation history of the galactic disk, therefore, intrinsically faint
(\& old) white dwarfs in the immediate solar neighborhood can be used
to determine an upper limit for the age of the galactic disk.

Most determinations of the white dwarf luminosity function have relied
on the use of Schmidt's $1/V_{\rm max}$ (Schmidt 1975) method for
magnitude and proper-motion selected samples, the behavior of which
has been demonstrated to follow a minimum variance maximum-likelihood
pattern for large samples. Additionally, recent numerical simulations
have also demonstrated that the $1/V_{\rm max}$ provides a reliable
estimator of the true LF, even in the case of small samples (Wood and
Oswalt 1998, and Garc\'{\i}a-Berro et al. 1999). However, the
conclusions from all these previous studies have been based on
noise-free data, where errors in the derived LF have been either
assumed to follow a Poisson distribution (valid only for large
samples), or where other simple estimates for the uncertainties have
been used.

In this paper we examine the faint-end ($M_V > +14$) behavior of the
disk white dwarf luminosity function using the $1/V_{\rm max}$ method,
but fully including the effects of realistic observational errors in
the derived luminosity function. We employ a Monte Carlo approach to
produce many different realizations of the luminosity function from a
given data set with pre-specified and reasonable errors in apparent
magnitude, proper-motions, parallaxes and bolometric
corrections. These realizations allow us to compute both a mean and an
expected range in the luminosity function that is compatible with the
observational errors.

We find that current state-of-the art observational errors, mostly in
the bolometric corrections and trigonometric parallaxes, play a major
role in obliterating (real or artificial) small scale fluctuations in
the luminosity function. We also find that a better estimator of the
true luminosity function seems to be the median over simulations,
rather than the mean. When using the latter, an age of 10~Gyr or older
can not be ruled out from the sample of Leggett, Ruiz, and Bergeron
(1998).

\end{abstract}


\section{Introduction} \label{intro}

The luminosity function (LF) for (disk) white dwarfs (WDs) is an
important observational tool to guide our understanding of the
evolution and fate of intermediate-mass systems, as well as for the
more general Galactic structure problems of the age of the Galactic
disk, as first pointed out by D'Antona and Mazzitelli (1978), and the
star-formation history of the solar neighborhood (Isern et
al. 1995). Indeed, WDs evolutionary time-scales represent a useful
tool for constraining the age of the disk of our Galaxy. The existence
of an abrupt fall-off in the observed WD~LF (see Liebert, Dahn and
Monet (1988), Leggett, Ruiz and Bergeron (1998)) has been interpreted
as an indication of the finite age of the Galactic disk (Winget et
al. 1987). By fitting the observations with theoretical WD~LFs, this
interpretation has been quantitatively explored by various
investigators (Wood 1992, and references therein). Not considered in
these studies has been an analysis of the effects of {\it
observational errors} on the {\it observationally derived}
LF. Recently, Wood and Oswalt (1998) and Garc\'{\i}a-Berro et
al. (1999), have performed a very comprehensive set of numerical
simulations studying the effects of different star-formation rates,
IMF, and kinematical prescriptions for (the progenitors of) WDs on the
derived {\it theoretical} WD~LF. However, all their simulations were
performed under the assumption that observational quantities were
noise-free.

In this paper we attempt to realistically quantify and characterize
the effects of observational errors on the derived LF for WDs by
comparing a LF derived assuming no errors at all, with LFs using
errors on the various basic observational parameters, applying the
$1/V_{\rm max}$ method.

In Section~2 we present a general discussion of the importance and
scope of determining the LF for WDs. In Section~3 a brief description
of the $1/V_{\rm max}$ method is given, while Section~4 describes
recent determinations of the WD~LF using this method. In Section~5 we
describe the basis of our numerical simulations. Sections~6 and~7
present the results of our simulations in terms of the global
uncertainties in the WD~LF using current data and the effects of
individual sources of errors, respectively. Section~8 outlines our
conclusions.

\section{The luminosity function} \label{luminosity}

The classical method for determining the LF of magnitude and
proper-motion selected samples is that proposed by Schmidt
(1975). This method, called the $1/V_{\rm max}$, stems from a
generalization of a method proposed earlier by Schmidt (1968) for
magnitude-limited samples. The method assumes that the LF does not
change (evolves) as a function of distance from the observer, and that
the sample is homogeneously distributed in space. The $1/V_{\rm max}$
method computes the LF by weighting the contribution of each observed
point by the equivalent volume where that particular object could have
been observed under the pre-specified survey constraints. Felten
(1976) has shown that the $1/V_{\rm max}$ method for magnitude-limited
samples is a minimum variance maximum-likelihood estimator and that,
for small absolute magnitude bins and very large samples ($>200$
objects per bin, not the case of WDs), it provides a reliable way of
estimating the parent (true) LF. Several modifications have been
proposed to the method in the case of magnitude-limited samples (Davis
and Huchra 1982, Eales 1993), most notably one that allows the
combination of different samples coherently (Avni and Bahcall
1980). However, the basic scheme to determine the LF of magnitude {\it
and} proper-motion selected samples has remained unchanged, and few
and limited numerical simulations have been carried-out to explore the
robustness and possible biases that the original method might have
when dealing with complete, but small, kinematically selected samples
(Wood and Oswalt 1998, Garc\'{\i}a-Berro et al. 1999).

Because the spatial density of WDs is rather small (about $3.4 \times
10^{-3}$~stars/pc$^3$ down to $M_V \sim +16.75$), it is important to
ensure that the method used to determine its LF is either free from
biases, or that they can be at least reliably corrected. Also, it is
important to understand the effects of the kinematic selection on the
resulting LF. For this purpose, Wood and Oswalt (1998) and
Garc\'{\i}a-Berro et al. (1999), have performed extensive numerical
simulations by creating fake catalogues of WDs in the solar
neighborhood from a pre-specified LF and a kinematical
description. Their (predicted) mock catalogues are an approximation to
true catalogues with similar selection biases in apparent magnitude
and total proper-motion. These mock catalogues are then passed to the
$1/V_{\rm max}$ from which a LF is predicted. This LF is then compared
to a range of input LFs, with different catalogue constraints, and
selection effects. The main results from the Wood and Oswalt (1998)
work are 1) the $1/V_{\rm max}$ method provides robust estimates of
the true local space density, 2) the age of the galactic disk must be
considered uncertain by about 15\% for the currently available sample
sizes, and 3) the bright-end of the derived LF shows substantial
deviations from the input functions, suggesting that is is difficult
to derive variations in the recent star formation history of the disk
from magnitude and proper-motion selected samples. Similarly,
Garc\'{\i}a-Berro et al. (1999) also find that 1) the simulated and
observed LFs are in excellent agreement, 2) the effect of a
scale-height law are important, specially at large intrinsic
luminosities (i.e., the bright end of the LF, which we do not consider
in this paper), 3) observational errors in the LF are well represented
by Poisson errors for samples of 200 stars or more, and 4) the
statistical uncertainty in the age of the disk is about 1~Gyr, in
agreement with the findings from Wood and Oswalt.

Unfortunately, the Wood and Oswalt's and Garc\'{\i}a-Berro's et
al. simulations have not included the effect of realistic
observational random errors in the key observational quantities, and
thus the effect of these errors on the resulting LF has not been
evaluated. This is precisely the motivation and scope of this work. Of
course, of particular interest, is the behavior of these simulations
with respect to the {\it slope} of the faint end of the WD~LF, which,
as mentioned in Section~(\ref{intro}), can be used as a constraint for
the age of the Galactic disk. Another point of interest is the level
at which the detailed shape (``wiggles'') on the WD~LF are real, for a
given sample-size, and can be interpreted as a consequence of the
evolution of WDs as a function of cooling age (Diaz-Pinto et al. 1994,
Hernanz et al. 1994).

\section{The $1/V_{\rm max}$ method}

The method proposed by Schmidt (1968, 1975) allows for a derivation of
the LF for a complete and spatially uniform sample of stars for which
we know their apparent magnitudes, parallaxes and (if used in the
sample selection), proper-motions. We also need to know the sample
selection (or survey) limits.

If we have a sample with a lower proper-motion limit $\mu_l$ and a
faint apparent magnitude limit $m_f$, the maximum distance $r_{\rm
max}$ over which any star can contribute to the sample is given by:

\begin{equation}
r_{\rm max} = min [p^{-1} (\mu/\mu_l) ; p^{-1} 10^{0.2(m-m_f)}] \label{rmax}
\end{equation}

where $p$ is the parallax, $\mu$ is the proper-motion, and m the
apparent magnitude.

Similarly, if the sample is only complete to an upper proper-motion
limit $\mu_u$ and a bright apparent magnitude $m_b$, the minimum
distance for inclusion would be:

\begin{equation}
r_{\rm min} = max[p^{-1} (\mu/\mu_u) ; p^{-1} 10^{0.2(m_b-m)}] \label{rmin}
\end{equation}

Finally, if the sample only covers a fraction $\beta$ of the sky, then
the maximum volume in which a star can contribute to the sample is:

\begin{equation}
V_{\rm max} = \frac{4}{3} \pi \beta (r_{\rm max}^3 - r_{\rm min}^3)
\end{equation}

The contribution to the LF from each star in the sample is then
$1/V_{\rm max}$, and the LF is calculated by adding the $1/V_{\rm
max}$ values over discrete magnitude intervals. For more details of
the method, the reader is referred to Schmidt (1968, 1975).

As can be seen from the above equations, the contribution from every
star in the sample to the overall LF is highly non-linear in terms of
the basic observational quantities, hence preventing an analytic
treatment of errors, specially in the case of small samples. This is
even more relevant if we consider that every observational point has
its own error budget, which highlights the need for doing full
numerical simulations of the effect of errors on the derived
LF. Indeed, Wood and Oswalt (1998) have pointed out that the noise
properties of the $1/V_{\rm max}$ are not well understood in the limit
of small samples (see also Felten 1976).

\section{Previous WD~LF determinations using the $1/V_{\rm max}$ method} \label{previous}

Liebert, Dahn \& Monet (1988, LDM88 thereafter) have presented
trigonometric parallaxes, optical colors, and spectrophotometric data
for a complete sample of intrinsically faint WDs (derived from the
Luyten Half Second catalogue, LHS, Luyten 1979) in the context of a
program to determine the faint end of the WD~LF. Using the classical
$1/V_{\rm max}$ method, they derived a LF which indicated a downturn
near $log(L/L_\odot) \sim -4.4$, a stellar density of $3 \times
10^{-3}$~stars pc$^{-3}$, and a derived age for the disk in the range
$7-10$~Gyr. More recently, Legget, Ruiz and Bergeron (1998, LRB98
thereafter) gathered new optical and near-IR data for the same cool
WDs in the LDM88 sample. Using stellar parameters derived from these
data and the more refined model atmospheres by Bergeron et al. (1995),
they re-derived the faint-end of the WD~LF, also using the $1/V_{\rm
max}$ method.  Comparing their LF with the (then) most recent cooling
sequences by Wood (1995), they derived a rather young age for the disk
of $8 \pm 1.5$~Gyr. In both cases, the uncertainty on the LF was
computed using the classical approach of assuming Poisson noise in the
counts of every absolute (or bolometric) magnitude bin, without
consideration of the actual observational errors for the quantities
involved in the LF determination. Therefore, only sampling errors were
considered, but not observational errors. In LRB98, furthermore, the
authors endeavored to estimate the uncertainties in the WD~LF by
calculating how many stars could be thrown into or out of each
magnitude bin due to errors in the bolometric correction - this is in
some sense a ``precursor'' of our numerical simulations (see
Sect.~\ref{numsim}.)

Figure~\ref{figcomp1} shows a comparison of the LF from LDM88 and
LRB98 as a function of absolute magnitude $M_V$, adopting a bin size
of 0.5~mag similar to that used by LDM88 and LRB98. Error bars are
Poisson bars, as adopted by these authors. We also indicate the LF as
a function of luminosity, using the same bolometric corrections (BCs
hereafter) adopted by the authors. In both cases (and in what follows
of our analysis) we have adopted a bright and faint apparent magnitude
limits of $V_b=+1$, $V_f= +19$, a lower and upper proper-motion limits
of $\mu_l= 0.8$~arcsec~yr$^{-1}$, $\mu_u= 10.0$~arcsec~yr$^{-1}$, and
a fraction of the sky covered of $\beta=0.5368$ (this last value is
derived from the fact that the LHS catalogue, on which the sample is
based, covers only the sky north of $\delta = -20^o$, and avoids the
Galactic plane). In the absolute magnitude range sampled by these WDs
the global normalization in the range $+12.75 \le M_V \le +16.75$ is
very similar, and equal to $\rho_* = 2.46 \times
10^{-3}$~stars~pc$^{-3}$ for LDM88 and $\rho_* = 2.54 \times
10^{-3}$~stars~pc$^{-3}$ for LRB98.

As we shall see, one of the key ingredients in determining a WD~LF
that could be compared with theoretically derived LFs is the
bolometric correction. To derive their observational WD~LF, LDM88 used
two extreme bolometric corrections, namely, no correction at all, and
another one based on the rather uncertain model atmospheres available
at that time. LRB98 on the other hand used not only the latest model
atmospheres, but they also fitted the detailed shape of the
theoretical spectrum to the observed optical and near-IR broad-band
colors for every single star in the sample, separately (details of the
fitting technique are given in Bergeron et al. 1997). In this last
case, errors in effective temperature were derived from uncertainties
in the fit, while the errors of the radius (surface gravity), where
derived by propagating the uncertainty in the trigonometric
parallaxes. The bolometric magnitude was then computed using $M_{\rm
bol} = -2.75 \log (L/L_{\odot}) + 4.75$, with $L= 4 \pi R^2 \sigma
T_{\rm eff}^4$.

\section{Numerical simulations} \label{numsim}

In this section we present the results from our numerical simulations,
fully including observational errors in all relevant quantities,
namely: apparent magnitude, bolometric corrections, proper-motion, and
parallaxes. It is assumed that quoted observational errors represent
the parent standard deviation, and that the true value follows a
Gaussian distribution function with the same standard deviation and
mean value as that given by the published data. This is probably an
idealization, but it should provide a better representation of the
data than just neglecting the observational errors, as it has been so
far the case. In every single realization of the LF (from now on
simply called a ``simulation'', and usually identified by a sequential
integer, or ``ID'' number), values with a mean and dispersion from
tabular input quantities taken either from LDM88 or LRB98 are randomly
drawn from a Gaussian distribution function for all the observables
(see, e.g., for the case of the simulated parallax,
Eq.~\ref{eqnerr}). The process of generating randomly Gaussian
distributed values for all the observables of a given data point is
repeated for all the data points in the input sample, thus creating a
single simulated sample. This simulated sample is then used to
construct the LF for that particular simulation using the $1/V_{\rm
max}$ method, and the whole process is repeated again to create a
different simulated sample and its respective LF. In the end, various
collective values, averaged over the simulations, are then
produced. In this way, it is possible to derive mean, median and
quartiles for the LF over a given set of simulations, as well as other
statistical indicators. For example, if $\Phi^i(M)$ is the luminosity
function at absolute magnitude $M$ for simulation ``$i$'', then the
mean-over-simulations LF is simply given by $<\Phi(M)> = \sum_{i=
1}^{N_{\rm simul}} \Phi^i(M) / N_{\rm simul}$, where $N_{\rm simul}$
is a (pre-specified) number of simulated LFs that have been generated
to create that mean LF. The associated mean stellar density in this
case would be given by $\rho_* = \int <\Phi(M)> \, dM $, or its
discrete summation counterpart (as we shall see in the next paragraph,
the stability of $\rho_*$ as a function of $N_{\rm simul}$ has been
used to define a lower limit for $N_{\rm simul}$ itself).

At the core of the simulation lies a (pseudo) random number
generator. We have tried two different generators in order to test the
sensitivity of our results to the adopted scheme, and found no
significant differences in the derived mean overall stellar density,
$\rho_*$, as a function of the number of simulations as long as the
number of simulations is larger than about 1,000 (see
Section~\ref{current}). In what follows we have therefore derived
collective values for the LF for 3,000 simulations but, evidently, our
results are independent of the total number of simulations above that
minimum number. We have found that 3,000 simulations is a good
compromise between stability of the simulations, RAM memory for array
storage, and CPU running time. The period (i.e., the number of calls
before producing correlations) of both random number generators tried
is, of course, much larger than the number of calls to the uniform
deviate routine (according to Press et al. (1997) the period is on the
order of $\sim 10^8$ for our adopted generator, see below). For
definiteness, we have finally adopted the routine ``ran1'' described
in the last edition of ``Numerical Recipes in Fortran'', from Press et
al. (1997). In order to avoid aliasing between different
``simulations'' (as defined in the previous paragraph), the ``seed''
for the random number generator is altered between successive
simulations. Note however that for a given simulation several calls to
the random number generator are required to produce uniform deviates
used in the error propagation (see Eqn.~(\ref{eqnerr})). Therefore,
the only purpose for updating the ``initial'' seed for a given
simulation is to render successive simulations as differentiated as
possible. Note also that after providing a seed for ``ran1'', the seed
gets altered by ``ran1'' itself, thus taking best advantage of the
large period in the generator.

In order to simulate true observational errors, we have adopted
Gaussian deviates derived using the routine ``gasdev'', also by Press
et al. The ``seed'' for the first call to ``gasdev'' is generated
congruently with the seed for that particular simulation, but altered
internally in ``gasdev'' for subsequent calls, as per the routine
described by Press et al. The value adopted for the observable is
simply given by, e.g., for the parallax:

\begin{equation}
p_{i,j} = p_j + \sigma_{{\rm p}_j} \times G_i \label{eqnerr}
\end{equation}

where $p_j$ is the (mean) observed parallax for star ``j'' in the
sample, with ``measurement error'' $\sigma_{{\rm p}_j}$, $G_i$ is the
Gaussian deviate (of zero mean and unity variance) for simulation
``i'', and $p_{i,j}$ is the i-th simulation value for the parallax of
star j. The same is performed for proper-motion, apparent magnitude,
and the bolometric correction (if necessary). The no-errors situation
is, of course, reproduced when all the $G_i$'s are set to zero.

A subtlety associated to the simulated sample values arises because of
the general survey restrictions $m_b \le m \le m_f$ and $\mu_l \le \mu
\le \mu_u$. In this case, if a simulated value falls outside the
survey limits because of under/over-shoot due to observational errors,
that object is eliminated from the sample, and its contribution to the
LF is suppressed.

\section{The current-sample WD~LF and its uncertainty} \label{current}

In this section we present the results from our numerical simulations,
using the simple technique described in the previous section. As
explained before, a ``convergence'' criterion for the simulations has
been used, by adopting the overall luminosity normalization (stellar
density) as the prime parameter. Of course, other criteria could be
used, but the basic point here is that the convergence criterion
ensures that the derived mean LF becomes {\it independent} of the
number of simulations. It is, therefore, the most representative value
for the LF given a sample and its errors.

Figure~\ref{figcomp2} shows the mean LF resultant from 3,000
simulations, using the errors quoted by LRB88 and LDM98
respectively. Since no errors for the proper-motion were given, it was
assumed a typical value of 10~mas~yr$^{-1}$ (where 1~mas=
1~milli-arcsec). As we shall see, the exact value for the
proper-motion error is less critical than uncertainties in the other
quantities, so this is probably a good estimate. The same
proper-motions were adopted for both studies. A few entries missing
errors for V magnitude in LDM88 were given a probable uncertainty of
0.05~mag, and the same magnitude errors were assumed for LRB98, which
is perhaps an overestimation of LRB98's photometric errors, but it is
not inconsistent with their statement that their ``photometric
uncertainties are 3\%''. The main differences between these two
studies comes from different parallaxes and their uncertainties, due
to improved parallax series using more plates, and the use of CCDs for
some of them, slightly different optical photometry, also improved by
the use of digital detectors, and different bolometric corrections and
estimated uncertainties coming from improved stellar interior and
atmospheric models and the addition of near-IR photometry used in the
LRB98 study. We have adopted a flat uncertainty of 0.14~mag in the BC
for the LDM88 sample; This is the value they quote for the difference
between two possible model BC corrections. This is probably an
overestimation of their true BC errors, which applies only to the
region of overlap where the comparison between different models was
done, but it provides an upper boundary to their BC errors. For the
LRB98 sample, we adopt their quoted BC uncertainties based mostly on
uncertainties in fitting their model atmospheres to the broad-band
optical and near-IR colors.

Figure~\ref{figcomp3} shows a comparison between the no-errors WD~LF
(from Fig.~(\ref{figcomp1})) and the LF derived using our full error
simulation (from Fig.~(\ref{figcomp2})). It is apparent that, while
the mean-over-simulation WD~LF is not so different (at least for the
brighter bins) in both cases, the error bars are quite a bit larger at
the fainter bin in the Monte-Carlo simulations. As a result, LRB98's
quoted disc age of 8~Gyr has to be taken with caution and, in fact,
their data does not rule out the possibility of an older disk, with an
age as large as 10~Gyr. It is also apparent from Fig.~\ref{figcomp3}
that the simulations indicate a rather long tail to fainter
luminosities in comparison with the abrupt decrease seen in the
no-error calculation. This has, of course, important implications for
the interpretation of the LF in terms of a well defined finite age for
the disk. Actually, as we shall see, this long tail is produced by a
few large-error excursions in the simulation leading to a biased mean
LF. Another important point is the exact break (if any!) in the
LF. This has been an outstanding issue over the years (c.f. the
extensive discussion of this on LDM88 or LRB98), the main difficulty
here coming from the fact that the position of the luminosity break
depends on the exact positioning of the bolometric magnitude bin,
which in the classical $1/V_{\rm max}$ method is fixed
arbitrarily. Both of these issues have motivated us to further explore
the behavior of the WD~LF, specially at the faint end.

We have devised a simple new version of the $1/V_{\rm max}$ method
which renders the resulting LF {\it independent} of the bin
positioning. For this purpose, we have adopted the same computational
scheme of the traditional Monte-Carlo method described above, but in
such a way that we have a ``moving box'' over absolute magnitude. The
main difference here is that in the classical method, the position of
the absolute (or bolometric) magnitude bins are pre-specified {\it a
priori}, whereas in this new moving box method, we only specify the
bright and faint absolute (or bolometric) magnitude limits, and an
arbitrarily large number of steps between them (so that the resulting
LF appears as a continuous function, rather than discrete as in the
classical case -an important feature if one is trying to look for
structure in the LF). Because we still need the absolute normalization
of the LF, the ``box'' has to be integrated over a pre-specified bin
width, but the {\it position} of the luminosity bins themselves can be
defined over an arbitrarily fine grid. Because the step between
successive boxes might be smaller than the bin width, errors from bin
to bin are not totally independent, and are thus highly
correlated. For this reason computing an error at each box position on
the absolute magnitude grid is less meaningful than for the case of
the traditional scheme (we can however, compute other statistical
properties of the LF, see below).

Figure~\ref{figcomp4} shows a comparison of the Monte-Carlo LF
computed on discrete intervals, and using our moving box, both for the
LRB98 data set. If, instead of using the mean LF, we adopt the median
over the simulations, we reproduce a sharp decline in the LF at the
faintest bins. Indeed, the quartiles, also shown on the figure,
indicate a rather tight distribution in comparison with the standard
deviation from the 3,000 simulations. This indicates, in turn, that
the distribution of predicted LF values, at a given luminosity, might
be quite skewed. If this is the case, then, extreme care has to be
taken when computing a ``mean'' LF considering the observational
errors: One must choose an indicator that resembles that of the most
representative LF value for a given observational data set.  From this
figure we can also see that the median LF compares very well with the
LF derived using no observational errors at all, except in that the
latter implies a fall-off at slightly brighter luminosities. The
extreme skewness of the LF is clearly demonstrated in
Fig.~\ref{lfdistrib} where the histogram of LF values is shown as a
function of bolometric magnitude. We see that, as we go to fainter and
fainter bins (solid points), the LF becomes more and more strongly
peaked at lower density values, with outliers to large stellar
density. This is mostly due to large-excursion outliers which turn
objects intrinsically dimmer in the simulations, and thus closer. As a
result, the volume sampled decreases, and the stellar density
increases. Another way of appreciating this effect, is to look at the
stellar density as a function of simulation. Figure~\ref{simulden}
shows the predicted overall stellar density as a function of
simulation, for the 3,000 simulations described above, and for the
LRB98 dataset. We see that, in comparison with the no-errors predicted
density, there is a small, but appreciable, over-density in the Monte
Carlo simulations, due to the sampled volume effect just mentioned. As
pointed-out before, the direct density leads to $\rho = 2.54 \times
10^{-3}$~stars/pc$^3$, whereas a simple fit to the data on
Fig.~\ref{simulden} indicates a stellar density of $\rho = (2.59 \pm
0.20) \times 10^{-3}$~stars/pc$^3$ (standard error).

We should emphasize that our derived WD~LF is still dependent upon bin
size. Indeed, the sample is still too small to be used as an effective
indicator of different star formation episodes in the disk. This is
clearly demonstrated in Fig.~\ref{lf_wigg}, where we have produced
three LFs by changing the bin size from 0.75~mag to 0.25~mag in steps
of 0.25~mag, and using a moving box with size half that of the bin
size, i.e., of the same order of the Nyquist frequency for the chosen
bin size (this was done to avoid strongly correlated errors from bin
to bin). As it can be seen from this figure, the exact position of the
sharp fall-off in the LF depends slightly on the bin width chosen,
and, furthermore, the ``wiggly'' features in the LF appear only for
the smallest bin size, indicative of the onset of large statistical
fluctuations due to the small samples concerned (even at relatively
bright magnitudes).

\section{The effect of observational errors on the WD~LF}\label{effects}

In this section we empirically explore the effects of different
observational error on the derived WD~LF. We start always from the same
sample (for definiteness the LRB98 dataset), but we fudge their errors
to different amounts in order to understand the behavior of the
$1/V_{\rm max}$ method, and the importance of the different sources of
errors, on the resultant LF.

We begin with a WD~LF derived using the moving box approach described
in the previous section. This would be the ``true'' WD~LF for this
data set (computed on a continuous set of bins), if there were no
observational errors. We then add, separately, errors in bolometric
corrections, magnitudes, parallax, and proper-motion, and compare
these LFs with that derived assuming all errors are zero. The outcome
of these simulations is a prescription, for observers (and theorists
as well!, see below), as to what parameters are more critical, and
should thus be refined further.

Figure~\ref{errorbudg_lrb} plots the quoted observational errors from
the LRB98 data as a function of apparent V magnitude. This plot gives
us an idea of the range of uncertainty in the observational
quantities. It is apparent that the uncertainty in bolometric
corrections (middle panel, $<\sigma_{\rm BC}> = 0.08 \pm 0.02$~mag) is
a lot larger than that of the direct photometry (upper panel,
$<\sigma_V>=0.028 \pm 0.004$~mag). This is an important point because
the WD~LF is equally sensitive to uncertainties in the BC and apparent
magnitudes. As for the parallax errors, we have $<\sigma_{\pi}>=4.3
\pm 0.7$~mas.

Figure~\ref{errorbudg1} shows, as a solid line, the (continuous) mean
WD~LF derived from the LRB98 data by assuming that the uncertainties
in all observables are zero (this LF ``looks'' different from the one
on Fig.~\ref{figcomp1}, which also assumes no errors, because the
later uses the classical $1/V_{\rm max}$ method, while in the former
we are using our ``moving box'' approach which produces a continuous
LF). We in turn start ``adding'' errors in various parameters, and
discussing their effect on the derived LF. In the simulations shown on
the previous sections, we have adopted a flat error for the
proper-motion of 10~mas~yr$^{-1}$. The exact value adopted is not
critical to the resultant LF. Indeed, a value 3 times as as large as
the assumed one does not produce any significant differences in the
derive LF. Only by the time the errors have gone up to as much as
100~mas~yr$^{-1}$ the LF starts showing the effects of these
errors. Furthermore, the effect only appears as a scale (or
normalization) factor in the overall LF (see upper panel on
Fig.~\ref{errorbudg1}), and not as a significant change of shape on
the LF, in marked contrast with the effect of errors on apparent
magnitude and bolometric corrections (see discussion in the next two
paragraphs, and the middle and lower panels of
Fig.~\ref{errorbudg1}). According to Dawson's analysis of the LHS
catalogue (Dawson 1986), an intercomparison of Luyten's proper-motions
and those derived from the much more accurate USNO parallax program
indicates that the rms error for a single star in the LHS catalogue is
29~mas~yr$^{-1}$, in agreement with Luyten's own estimates. From about
50\% of the sample published so far by Wroblewski \& Torres (1990, and
subsequent papers), Costa (2000) finds a slightly larger dispersion of
38~mas~yr$^{-1}$ (1,262 stars) between their proper-motions and those
from Luyten for his LTT sample (although there is a contribution to
this dispersion from their own measurement errors which is in the
range 5-25~mas~yr$^{-1}$), while Ruiz et al. (2000) obtains a
dispersion of 36~mas~yr$^{-1}$ (23 stars, with similar internal errors
as those of the Costa's sample) as derived from their respective
proper-motion surveys in selected areas of the southern sky, which has
recovered many (and added a few new) of the large proper-motions stars
from the LHS catalogue. We thus conclude from these simulations that
the current proper-motion uncertainties do not contribute
significantly to the overall LF error.

Bolometric corrections are, of course, a critical step when deriving a
WD~LF that can be compared with theoretical models. Unfortunately,
errors in this parameter are still rather large (see middle panel on
Fig.~\ref{errorbudg_lrb}), and do produce a large negative impact on
the accuracy of the derived LF. The middle panel on
Fig.~\ref{errorbudg1} shows the effects on the WD~LF of uncertainties
amounting to 0.05~mag (red line) and 0.10~mag (green line) on the BC.

Uncertainties in apparent magnitude have the same effect on the LF as
do uncertainties in the BC, but their observational errors are a lot
smaller, and therefore do not play an important role in the final
WD~LF (see upper panel on Fig.~\ref{errorbudg_lrb}). Parallax
uncertainties also turn out to be relevant. The bottom panel on
Fig.~\ref{errorbudg1} shows the effects of 1~mas (red line), and 3~mas
(green line) parallax errors on the derived LF. From these plots we
can clearly see that, for a ``typical'' uncertainty of $\sigma_{\rm
BC} \sim 0.1$~mag and $\sigma_{\pi}=4$~mas on these two quantities,
the contribution to the ``smearing'' of the WD~LF is, at present,
equally represented by errors on the trigonometric parallaxes and
bolometric corrections (all the blue lines on Fig.~(\ref{errorbudg1})
indicate the contribution of the true errors from proper-motions, BC
and parallaxes, respectively, to the resultant LF).

The explanation for the markedly different behavior of the derived
WD~LF upon errors on proper-motion {\it vs.} errors in bolometric
corrections and parallaxes is actually easy to understand: As it can
been seen from Eqns.~(1) through (3), the contribution to the
luminosity function depends on proper-motions, parallaxes, and
apparent magnitudes, whereas the placement in luminosity of a given
observed data point depends {\it exclusively} on the bolometric
correction, apparent magnitude and (the logarithm of) the parallax,
but is independent of the object's proper-motion. Therefore, as stated
before, errors in proper-motion will only impact upon the scale of the
LF, and will not displace points in luminosity, while uncertainties in
the other observables would impact both, the LF normalization, and the
actual luminosity where that object is contributing to the overall LF.

In terms of the extent to which the sample analyzed is complete, our
simulations indicate that, with the adopted survey boundary
constraints, it is probably incomplete. Figure~(\ref{complete}) (upper
panel) shows the classical estimator $<V/V_{\rm max}>$ (which should
approach the value 0.5 for a complete sample) as a function of the
simulation ID for the LRB98 sample. As it can be seen, no simulation
brings the above value closer to 0.5. Indeed, for large-excursion
errors in which one or two objects fall outside the survey boundary,
the value of $<V/V_{\rm max}>$ decreases, leading to two parallel
sequences to the main set of values. The mean error on $<V/V_{\rm
max}>$ remains, however, constant and quite small (lower panel on
Fig.~(\ref{complete})). We can easily explore the effect on the
derived LF and sample incompleteness due to potentially erroneous
survey boundaries with the aid of our simulations. In
Fig.~(\ref{surboun}) we show the effect of introducing variations in
the survey limiting magnitudes, and in the proper-motion limits. As
mentioned in Section~\ref{previous} we have adopted $V_b=+1$, $V_f=
+19$, $\mu_l= 0.8$~arcsec~yr$^{-1}$, $\mu_u=
10.0$~arcsec~yr$^{-1}$. We find that the derived LF is insensitive to
the value of $V_b$, and that even by adopting the extreme case
$V_b=+5$ the LF does not change at all. Also, surprisingly, by
adopting a very conservative cut $V_f= +18$, the LF is only altered
mildly. Proper-motions do have, however, an important role in the
sample selection, and in the resultant LF (as found also by the
simulations from Wood and Oswalt (1998) and Garc\'{\i}a-Berro et
al. (1999)). While a cut at $\mu_u= 5.0$~arcsec~yr$^{-1}$ does not
change appreciably the LF, a lower value of $\mu_u=
2.0$~arcsec~yr$^{-1}$ does change the shape of the LF appreciably. The
only remaining source of uncertainty, i.e., the lower proper-motion
limit, does also seem to have a big influence on the derived LF but
only for $M_{bol} \le +15.0$ (see Fig.~(\ref{surboun})), and only when
adopting a very conservative $\mu_l=
1.0$~arcsec~yr$^{-1}$. Figure~(\ref{surboun}) also shows the
encouraging news that the exact break at the faint end of the WD~LF is
not extremely sensitive to the survey boundary and/or incompleteness
effects. From a comparison of Luyten's catalogued stars and their
newly discovered large proper-motion stars in selected areas of the
Southern sky, Ruiz et al. (2000) find that the LHS catalogue might be
actually incomplete in a more severe way than previously thought, at
least in the Southern sky. For example, while they corroborate that
the LHS catalogue is incomplete for $\mu \le 0.8$~arcsec~yr$^{-1}$,
they also find that incompleteness sets for $m_{\rm R} > 14$, i.e.,
several magnitudes brighter than claimed by Luyten, although the
overall number of stars involved in the comparison is small ($\sim
50$). This might not necessarily apply directly to the Northern sky
sample analyzed here, since it is known that the Southern plates used
by Luyten (those of the Bruce proper-motion survey) were shallower
($m_{\rm pg_{lim}} \sim 15.5 - 16.0$) than its Northern counterpart
(from the Palomar proper-motion survey, with $m_{\rm pg_{lim}} \sim
21.2$). One must note that the completeness test for the noise-free
classical case also shows signs of incompleteness, having a value of
$<V/V_{\rm max}> = 0.367 \pm 0.046$ for the LRB98 dataset (hence the
suggestion of a significant incompleteness or erroneous survey
boundaries is nothing particular to our simulations!).

From the preceding simulations, it is interesting to notice that the
only way we can increase the value of $<V/V_{\rm max}>$ is by adopting
a smaller $\mu_u$. A value of $\mu_u= 2.0$~arcsec~yr$^{-1}$ implies
$<V/V_{\rm max}> = 0.447 \pm 0.051$, whereas $\mu_u=
1.5$~arcsec~yr$^{-1}$ leads $<V/V_{\rm max}> = 0.594 \pm 0.062$ (see
Fig.~(\ref{sup_complete})). By comparison, the adopted standard survey
boundaries have $<V/V_{\rm max}> = 0.366 \pm 0.046$. If the
incompleteness is due to a poor definition of the survey boundaries,
then this suggests that Luyten might have actually missed some of the
large ($\mu \ge 1.5-2.0$~arcsec~yr$^{-1}$) proper-motion stars, a fact
also stated by Dawson (1986) in connection with the luminosity
function of halo stars, as derived from the LHS catalogue. Our
simulations also imply that he only needed to have missed stars moving
with proper-motions larger than 1.5 to 2.0~arcsec~yr$^{-1}$, but
smaller than 4.0 to 5.0~arcsec~yr$^{-1}$, since the derived LF (and
the corresponding completeness test) is insensitive to $\mu_u$ for
$\mu_u \ge 5.0$~arcsec~yr$^{-1}$. Even though the LF is influenced
heavily for $M_{bol} \le +15.0$ by adopting $\mu_l=
1.0$~arcsec~yr$^{-1}$, $<V/V_{\rm max}>$ does not change substantially
in this case ($0.372 \pm 0.050$), indicating that the source of
incompleteness is due to, both, faint and large proper-motion stars
missing from the current sample.

\section{Discussion and Conclusions}

As was explained towards the end of Sect.~\ref{numsim}, simulated
sample points whose observables fall outside the survey limits are
eliminated from the overall sample. This procedure might be called
into question because the resulting WD~LF has a different number of
stars than the number contained in either the LDM88 or LRB98
samples. Indeed, this strategy can only {\it reduce} the number of
objects in the realization, whereas in the real world, objects can be
{\it added} to the sample as well, since the same observational errors
will occasionally add stars to the sample that were originally outside
the survey limits. The non-conservation of sample data points is
clearly seen in the second, parallel sequence in $<V/V_{\rm max}>$
that appears in Figs.\ref{complete} and \ref{sup_complete}. The
question is, then, to what extent the simulations displayed in Fig. 2,
which show a long tail of intrinsically faint stars with small implied
observational errors, might be an artifact of the non-conservation of the
total numbers of stars in the (simulated) sample?

In Sect.~\ref{effects} we have interpreted the small value for
$<V/V_{\rm max}>$ (see also Fig.\ref{complete}) as a reflection of an
incorrect definition of the survey limits, and we point out that the
second sequence in $<V/V_{\rm max}>$ is due to large-excursion errors
in which one or two objects fall outside the adopted survey
boundary. This is the same explanation we advanced for the faint tail
of the WD~LF in Sect.~\ref{current}. But, one might also question this
result, and wonder whether the fact that this quantity departs from
the expected value of 0.5 may not be a reflection of an incorrect
treatment of these large-excursion errors, as suggested in the
previous paragraph, rather than reflecting an incorrect treatment of
the survey limits.

To elucidate these important questions we run a few more simulations
where we discarded the entire realization of the WD~LF if the number
of objects was not conserved, and proceeded to the next
realization. In this case the simulated samples were, by construction,
always {\it similar} to the input sample, save for the exact values of
the observables which departed form the input values by an amount
specified only by the adopted errors, but still within the adopted
survey boundaries. The results of these simulations are shown in
Fig.~\ref{fig_ref1} (upper panel), which compares the LF derived in
the case where the sample is not conserved (i.e., the same LF as for
the LRB98 dataset shown in Fig.~\ref{figcomp2}, derived from 3,000
simulations) with the LF obtained in the case when the sample is
conserved (2,557 simulations out of 3,000 initial simulations, i.e.,
some 15\% of the simulations lost to extremely high-residual
excursions). As it is obvious from this figure, the resultant LFs are
quite similar, and both exhibit the same tail to faint
luminosities. This implies that the large-excursion simulated
observables responsible for the faint tail of the LF are not large
enough to alter the input sample significantly (therefore avoiding the
problems mentioned in the first paragraph above), while having an
important effect on the faint portion of the LF itself due to the
highly skewed distribution of LF values at a given luminosity shown in
Fig.~\ref{lfdistrib}. The overall completeness factors do not change
either, having in both cases the value $<V/V_{\rm max}> = 0.366 \pm
0.046$ (and quite close to the no-error case with $<V/V_{\rm max}> =
0.367 \pm 0.046$), thus alleviating the concerns expressed in the
previous paragraph.

In light of the above results, it is interesting to explore in some
detail the selection effects acting upon the derived simulated LFs and
inclusion/exclusion of objects near the survey limits. For this
purpose, we calculated the number of times that the selection effect
was either proper-motion or apparent magnitude, according to the first
and second terms respectively of Eqs.~\ref{rmax} and~\ref{rmin}, as a
function of bolometric magnitude. For the bolometric WD~LF exhibited
in the upper panel of Fig.~\ref{fig_ref1}, $r_{\rm min}$ was {\it
always} determined by the proper-motion, while the value of the
$r_{\rm max}$ was determined by a mixture of both proper-motion and
apparent magnitude. For $N_{\rm obj}$ objects in the input sample, and
$N_{\rm simul}$ simulations, the maximum number of times that a
criteria could be used is just $N_{\rm obj} \times N_{\rm simul}$. In
the lower panel of Fig.~\ref{fig_ref1} we show the histogram of
selection criteria for $r_{\rm max}$ for 43 objects (LRB98 dataset),
and 2,557 effective simulations (keeping the number of objects fixed,
see previous paragraph). As it can be seen from the figure, in the
vast majority of the simulations, the primary selection criteria in
the whole range $13 < M_{\rm bol} \le 17.5$ is determined by the
object's proper-motion, and {\it not} by its apparent magnitude, a
result already found by Wood and Oswalt (1998) and Garc\'{\i}a-Berro
et al. (1999, see related discussion below). Additionally, it was
demonstrated in Sect.~\ref{effects} and Fig.~\ref{errorbudg1} (top
panel), that errors in proper-motion will only impact upon the scale
of the LF, but will not displace points in luminosity, while
uncertainties in the other observables (in particular apparent
magnitude and bolometric corrections) will impact both, the LF
normalization, {\it and} the actual luminosity where that object is
contributing to the overall LF. We therefore can reasonably suggest
that the true significance of the by and large proper-motion selection
criteria at faint magnitudes, intermingled with a much less often
magnitude selection criteria, is responsible for keeping the number of
objects fixed within the survey boundaries (proper-motion selection,
changes LF normalization but not the object's luminosity) while
producing a long tail to faint magnitudes (magnitude selection,
changes normalization and luminosity, but only in a small fraction of
the simulations, see Fig.~\ref{fig_ref1}).

We can therefore conclude that our simulations indicate that LRB98's
data, when properly accounting for observational errors, does not rule
out a disk with an age as large as 10~Gyr. This is good news because
previous studies that find ages of 8~Gyr or younger using similar
datasets are difficult to reconcile with an halo age of 15~Gyr
(inferred from old globular clusters) given that Galactic formation
and chemical evolution models suggest a delay of, at most, 3~Gyr
between the onset of star formation in the halo and in the local disk
(Wood and Oswalt 1998). On the other hand, a value as large as 13~Gyr,
found by Garc\'{\i}a-Berro et al. (1999), can probably be ruled
out. Garc\'{\i}a-Berro et al. attribute this large derived age to the
effects of the scale height, but we should note that our study is
restricted to the faintest portion of the LF (with $r_{\rm max}$ from
Eq.~({\ref{rmax}) always smaller than 50~pc, or about 1/3$^{\rm rd}$
of a scale height), where the effect of the WD scale-height is
irrelevant because of the local nature of the sample. We note that
Galactic open clusters can also set constraints on the age of the
disk. NGC~6791 is the oldest known open cluster with a well-determined
age, in the range 7~to 10~Gyr (Tripicco et al. 1995), albeit Scott el
al. (1995) find that the kinematics (space velocity) for this object
is somewhat peculiar. The extreme 12~Gyr age of Berkeley~17, believed
to be one of the oldest disk clusters (Phelps 1997), has been recently
revised by Carraro et al. (1999) using near-IR photometry, leading to
a younger age of 8-9~Gyr. Also, Jimenez et al. (1998), using Hipparcos
data, have found an upper limit for the age for the disk field
population in the solar neighborhood of $11 \pm 1$~Gyr (see also
Bertelli et al. 1999), which would be in agreement with our revised
(older) age from the WD~LF. Also, we find that current observational
uncertainties and sample sizes do not allow us to establish the
existence of small scale features in the WD~LF which could be
indicative of different episodes of star formation in the disk. This
could only be alleviated by dramatically increasing the currently
small samples, as also emphasized by the simulations performed by Wood
and Oswalt (1998), and Garc\'{\i}a-Berro et al. (1999).

Both Wood and Oswalt (1998) and Garc\'{\i}a-Berro et al. (1999) have
found that the primary selection criteria at low luminosities is the
proper-motion. In our simulations we see a related effect, where a
larger proper-motion uncertainty (which affects the selection
criteria) leads to a change of the normalization of the LF, whereas
its overall shape does not change dramatically (see
Fig~\ref{errorbudg1}, upper panel). By using the $<V/V_{\rm max}>$
completeness criteria, we also find that the LRB98 sample seems to be
missing faint ($M_{bol} > +15.0$), large proper-motion ($\mu >
2.0$~arcsec~yr$^{-1}$) stars, and that the sample is only complete for
$\mu \le 1.5-2.0 $~arcsec~yr$^{-1}$. However, we find that the precise
luminosity break at the faint end of the WD~LF is not extremely
sensitive to the survey boundary and/or incompleteness effects (see
Figure~(\ref{surboun})).

In summary, we have found that most of the current uncertainties in
the observational WD~LF come from uncertainties in bolometric
corrections and in parallaxes, while photometry and proper-motions
play a minor role. Although this effect might be captured in
Fig.~\ref{errorbudg_lrb}, which simply displays the distribution of
errors in the bolometric corrections and parallaxes (and which, of
course, does not require any of the statistical discussion in the rest
of the paper), the impact of these uncertainties upon the Monte-Carlo
derived WD~LF for 3,000 simulations is fully shown in
Fig.~\ref{errorbudg1}. This last figure clearly shows that,
refinements on theoretical models (such that $\sigma_{\rm BC} \le
0.05$~mag) and parallaxes (with $\sigma_{\pi} \le 1$~mas), as well as
larger samples ($N_{\rm samp} \sim 200$, see Wood and Oswalt 1998),
should be primary goals in order to produce a better luminosity
function for white dwarfs.

Our Monte-Carlo simulations using the $1/V_{\rm max}$ method can be
recreated for any other proper-motion and/or magnitude selected
samples - a simple ASCII input table with mean values and probable
errors is all what it is needed. All the software and help on how to
use it is available from RAM.

\acknowledgments

RAM acknowledges useful comments from Dr. Gaspar Galaz (OCIW) and from
the late Dr. Olin Eggen (CTIO) and Prof. Claudio Anguita
(U.de~Ch.). RAM also acknowledges the continuous support from a
C\'atedra Presidencial en Ciencias to Dr. M.T. Ruiz. MTR received
partial support from C\'atedra Presidencial en Ciencias and Fondecyt
grant \#~1980659. We thank Dr. Leandro G. Althaus (UNLP) for sending
us their theoretical LFs. Finally, an anonymous referee made important
comments regarding the validity of our faint-luminosity tail on the
simulated LFs, to him we owe the relevant discussion centered around
Fig.~\ref{fig_ref1}.

\newpage

\figcaption[mendezr.fig1.ps]{A comparison of the LDM88 (solid circles)
and LRB98 (solid squares) WD~LF. A bin width of 0.5~mag, and the same
bin centering adopted by LDM88 and LRB98 have been used. A shift of
+0.1~mag in the magnitudes has been applied to the LRB98 data in order
to avoid crowding. The upper panel is for $M_V$, while the lower panel
is for $M_{bol}$, adopting the bolometric corrections applied by LDM88
and LRB98 respectively. Errors bars indicate Poisson error
exclusively. This figure shows, basically, our advance in 10~years of
optical \& IR photometry, parallaxes, bolometric corrections and
interior physics and stellar atmosphere modeling for these stars, as
reflected upon the LF. \label{figcomp1}}

\figcaption[mendezr.fig2.ps]{Similar to Figure~\ref{figcomp1}, but for
the mean-over-simulations WD~LF for 3,000 Monte-Carlo simulations
adopting the errors quoted by LDM88 and LRB98. By comparing with
Fig.~\ref{figcomp1}, it is apparent that the Monte-Carlo errors are
smaller than the Poisson errors at brighter magnitudes where the
sample is larger and less subject to observational errors. However, we
can also see that Poisson errors {\it underestimate} the true expected
LF uncertainties at the fainter bins. The mean value for the LF are,
however, quite similar in both cases. \label{figcomp2}}

\figcaption[mendezr.fig3.ps]{Bolometric WD~LF. Similar symbols as for
Fig.~\ref{figcomp1}. In this case, LRB98 has been shifted by +0.04 in
$\log (L/L_{\sun})$ in order to avoid crowding. The upper panel is for
no-observational errors, while the lower panel shows the results for
the mean LF after 3,000 simulations. The lines indicates the latest
theoretical WD~LF published by Benvenuto and Althaus (1999), based on
carbon-oxygen core WDs, {\it including} the release of latent heat
during crystallization (Salpeter IMF, constant star-formation rate
over the age of the disk). The dotted line is for a 6~Gyr disk, while
the dot-dashed is for a 10~Gyr disk. Theoretical LFs have been
normalized to a total $\rho = 3.39 \times 10^{-3}$~stars/pc$^3$. The
upper panel shows that, as found by LRB98, the WD~LF is not
inconsistent with a disk age of about 8~Gyr. However, in the lower
panel, our Monte-Carlo approach indicates that the large error-bars at
the faint end {\it could} allow for an older disk given current
observational uncertainties and sample sizes. \label{figcomp3}}

\figcaption[mendezr.fig4.ps]{Bolometric LF for the LRB98 dataset. The
solid squares with error bars indicate the LF using the Monte-Carlo
mean LF on discrete 0.5~mag bin intervals (same as in
Fig.~\ref{figcomp3}, lower panel), while the open squares reproduces
the LF derived in the case of no errors, shifted by +0.04 (same as the
solid squares in Fig.~\ref{figcomp1}). The solid line shows the mean
LF using our moving-box approach. In this case, the LF has also been
integrated over a 0.5~mag bin. The dashed line indicate the median
over simulations LF from the very same simulation that generated the
plotted mean LF, while the dotted lines indicate the lower 25\%
quartile and the upper 75\% quartile on the distribution of LF values
as a function of luminosity. The big difference between the mean and
median LF at faint magnitudes indicates a highly skewed distribution
of LF values, as it is indeed found (see
Fig.~\ref{lfdistrib}). Surprisingly, we can also see that the median
LF approaches very well the LF derived in the case of no
errors. \label{figcomp4}}

\figcaption[mendezr.fig5.ps]{Distribution of Monte Carlo LF values as
a function of bolometric luminosity for the LRB98 data set in 0.5~mag
bins. The plot shows the evolution of the histogram of LF values as a
function of bolometric magnitude. The LF has been derived from 3,000
simulations leading to the mean and median LF for LRB98 shown in the
previous figures. The symbols are, in decreasing luminosity, as
follows: Open circles for $M_{\rm bol} = +14.5$, open squares for
$M_{\rm bol} = +15.0$, open triangles for $M_{\rm bol} = +15.5$, open
stars for $M_{\rm bol} = +16.0$, filled circle for $M_{\rm bol} =
+16.5$, filled squares for $M_{\rm bol} = +17.0$, and filled triangles
for $M_{\rm bol} = +17.5$. \label{lfdistrib}}

\figcaption[mendezr.fig6.ps]{Predicted WD space density as a function
of simulation, for the 3,000 simulations leading to the LF on the
previous figures (dots). The horizontal solid line indicates the
density derived in the classical no-errors method. Changes in the
observables allowed within their uncertainties lead to a
``spill-over'' to adjacent luminosity bins that increase the effective
density by a small fraction due to a volume sampling effect (see
text). Although there is a systematic shift, the dispersion across
simulations is large enough that the Monte-Carlo mean density and the
direct density are {\it almost} the same within the dispersion of the
former. \label{simulden}}

\figcaption[mendezr.fig7.ps]{Continuous median WD~LF from the LRB98
data set adopting three different bin widths of 0.75~mag (dashed
line), 0.5~mag (solid line, adopted throughout this paper), and
0.25~mag (dot-dashed line). As it can be seen, rapid changes in the LF
appear only for the smallest bin width due to the small number of
objects per bin. At the faint end, the small bin-width effect is seen
as an uncertainty in the exact fall-off luminosity.\label{lf_wigg}}

\figcaption[mendezr.fig8.ps]{Quoted observational errors (dots) as a
function of apparent V magnitude for the LRB98 sample. Proper-motion
errors have been assumed to be 10~mas (see text), and are thus not
plotted. The upper, middle and lower panels are for uncertainties in
apparent magnitude, BCs, and trigonometric parallaxes
respectively. The horizontal line indicates the straight mean error
for all quantities. \label{errorbudg_lrb}}

\figcaption[mendezr.fig9.ps]{Linear continuous (using our moving box
approach) WD~LF as a function of uncertainties in proper-motion (upper
panel), BC (middle panel), and parallax (lower panel). In all panels,
the black solid line is for no errors while the blue solid line is for
the quoted (true) errors. In the upper panel the red line is for a
proper-motion uncertainty of 30~mas~yr$^{-1}$ (more representative of
the LHS catalogue), while the green line is for an extreme error of
100~mas~yr$^{-1}$, larger by a factor of three than the expected
errors in the LHS catalogue. As it can be seen from the plot,
proper-motions are not a significant source of uncertainty on the
derived LF --- they only affect the bin-to-bin normalization of the
LF, but do not broaden the luminosity distribution. In the middle
panel, the red line is for an uncertainty of $\sigma_{\rm BC} =
0.05$~mag, while the green line is for an uncertainty of $\sigma_{\rm
BC} = 0.10$~mag. In the lower panel, the red line is for a parallax
uncertainty of only 1~mas, while the green line is for an uncertainty
of 3~mas. The largest source of uncertainty in the present WD~LF is
found to come from uncertainties in both, the bolometric corrections
and the trigonometric parallaxes. \label{errorbudg1}}

\figcaption[mendezr.fig10.ps]{Completeness fraction (top panel),
measured in terms of $<V/V_{\rm max}>$ as a function of simulation,
for the 3,000 simulations resulting in the bolometric LF of
Fig.~\ref{figcomp2}. The lower panel shows the expected mean error in
$<V/V_{\rm max}>$ for the respective simulations. The horizontal line
indicates the values for the classical error-free case with $<V/V_{\rm
max}> = 0.367 \pm 0.046$. A truly complete sample should have
$<V/V_{\rm max}> \sim 0.5$ (Schmidt 1968,1975, Felten 1976) whereas
the LRB98 dataset has $<V/V_{\rm max}> \le 0.38$, indicating that
either the sample is somewhat incomplete, or the survey boundaries are
erroneous. \label{complete}}

\figcaption[mendezr.fig11.ps]{Linear continuous median WD~LF for
different survey boundaries. The solid line is for the standard survey
boundaries (dashed line on Fig.~\ref{figcomp4}), the dashed line for
$\mu_l = 1.0$~arcsec~yr$^{-1}$, the dot-dashed line for $\mu_u =
2.0$~arcsec~yr$^{-1}$, and the dotted line is for the extreme case
$\mu_u = 1.5$~arcsec~yr$^{-1}$. As can be shown, the lower
proper-motion limit does not have any effect on the fall-off
luminosity, but it has a great influence on the LF for $M_{bol} \le
+15.0$. The large proper-motion cut, however, influences both the
bright \& faint portions of the LF and, in addition, changes the
completeness of the sample (see Fig.~\ref{sup_complete}.)
\label{surboun}}

\figcaption[mendezr.fig12.ps]{Similar to Fig.~(\ref{complete}), but
with different values for the survey boundaries (see text). Green dots
are for $\mu_u =2.0$~arcsec~yr$^{-1}$ (with $<V/V_{\rm max}> = 0.447
\pm 0.051$), while red dots are for $\mu_u =1.5$~arcsec~yr$^{-1}$
(with $<V/V_{\rm max}> = 0.594 \pm 0.062$). This suggests that the
data is complete ( $<V/V_{\rm max}> = 0.5$) only for $\mu_u \ge
1.7$~arcsec~yr$^{-1}$. \label{sup_complete}}

\figcaption[mendezr.fig13.ps]{Top panel: Mean discrete bolometric
WD~LF for the LRB98 dataset for 3,000 Monte-Carlo simulations
discarding individual simulated objects as their errors move them out
of the survey boundaries (solid squares - same LF as solid squares in
lower panel of Fig.~\ref{figcomp2}), or by discarding the simulated LF
altogether from the computation of the mean LF if the simulated value
falls off the adopted boundaries (open squares, 2,557 effective
simulations). As it can be readily seen, the long tail to faint
luminosities remains intact even if we exclude the influence of
extreme simulated outliers while keeping the sample size fixed. Lower
panel: Log of the number of times that either a proper-motion (solid
line) or apparent magnitude (dashed line) criteria is adopted to
calculate $r_{\rm max}$ from Eq.~\ref{rmax} to compute the (open
squares) LF displayed in the top panel. As it is obvious, the {\it
primary} selection effect at faint magnitudes is proper-motion (see
text for significance of this). \label{fig_ref1}}

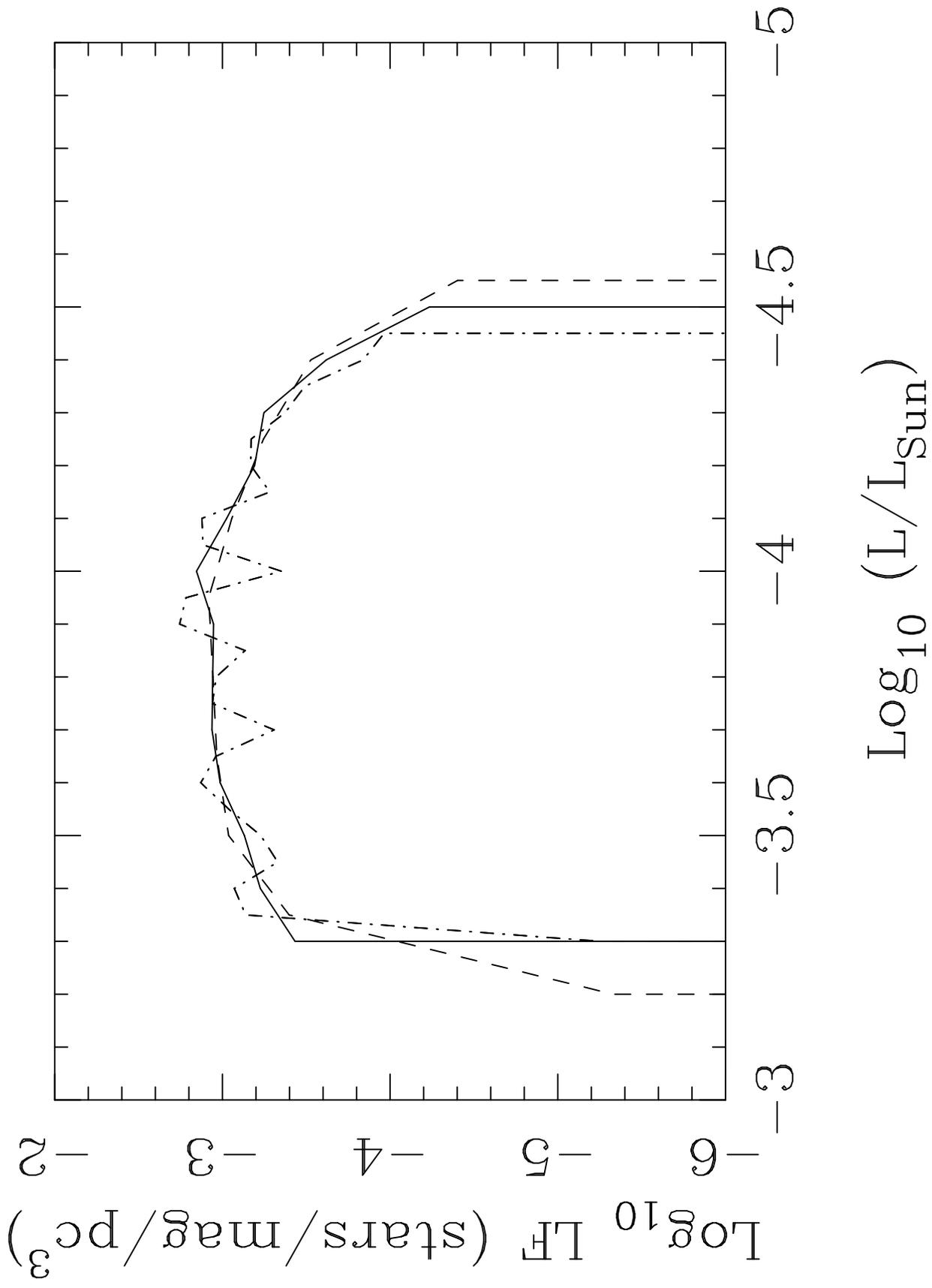
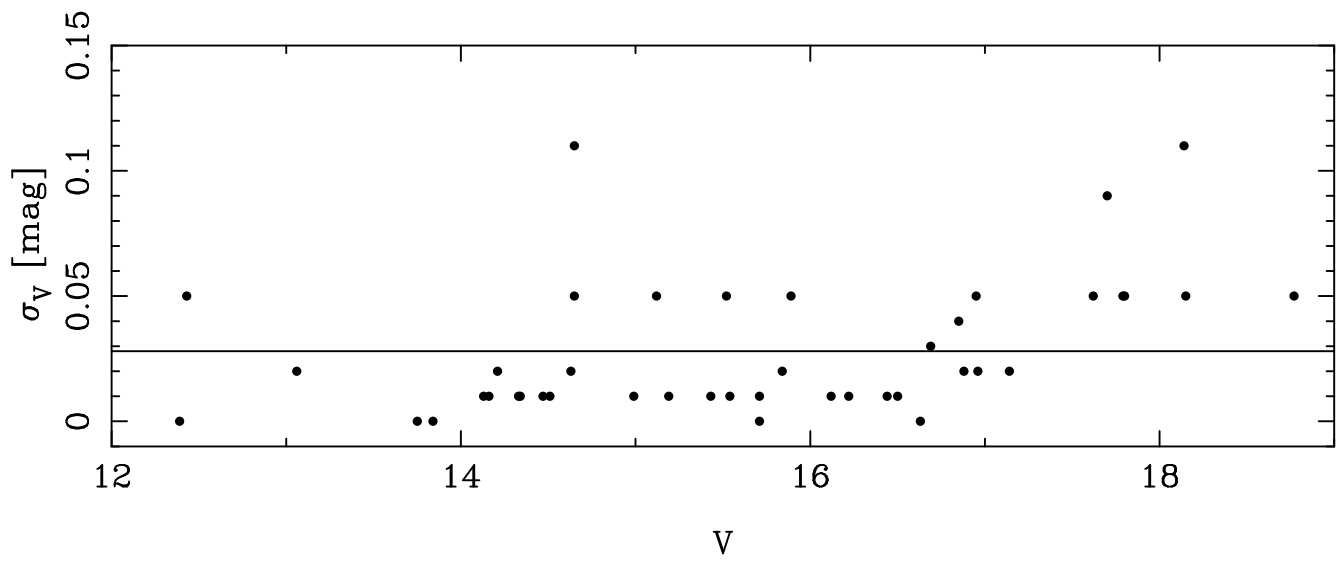
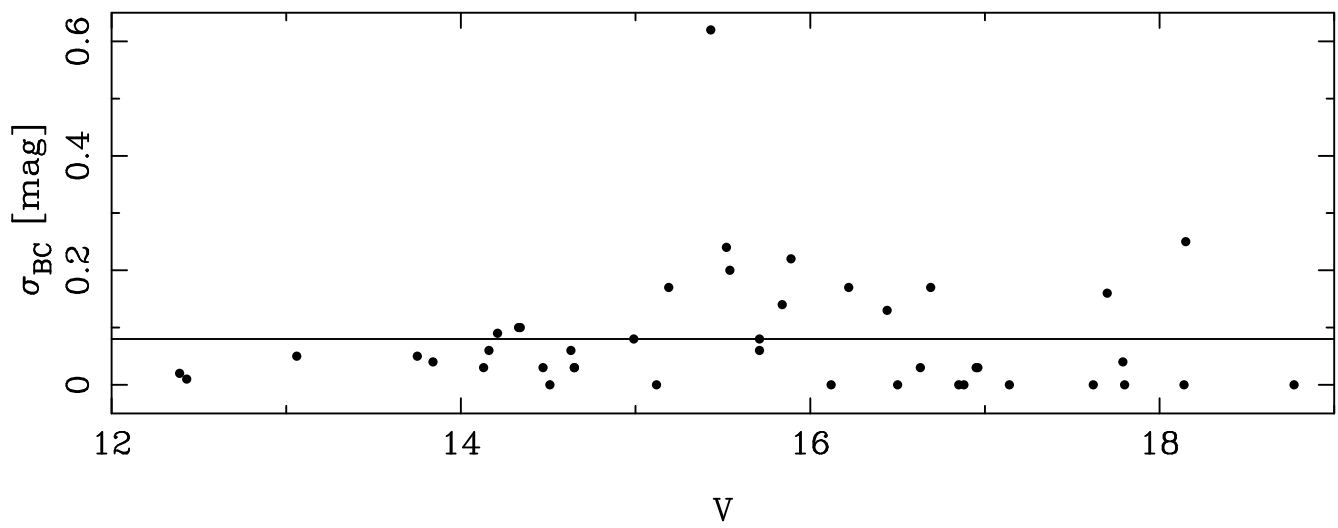
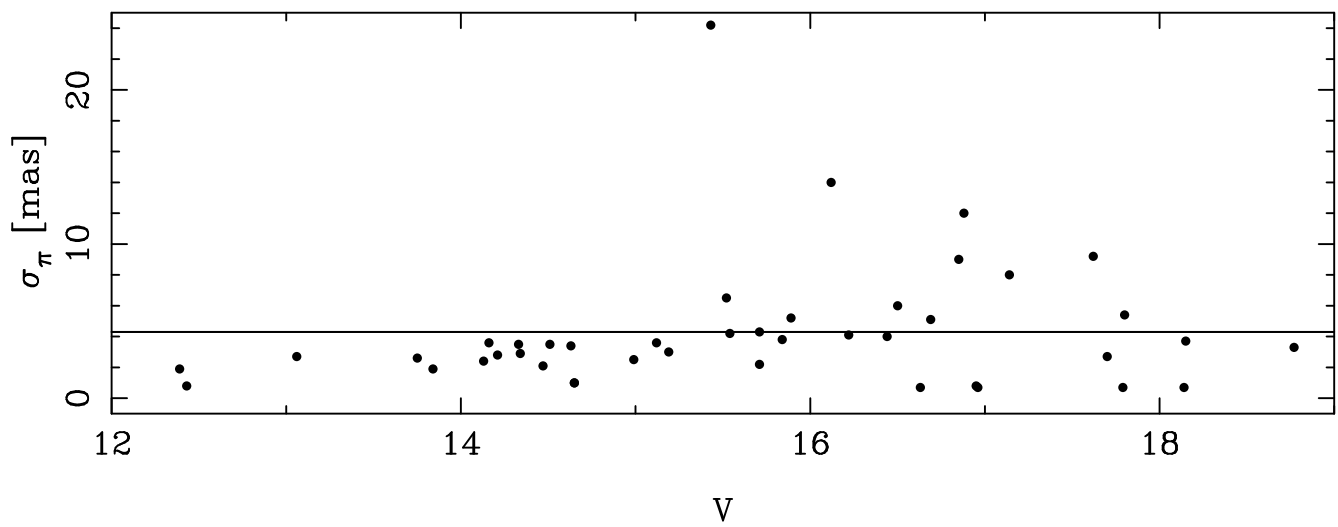
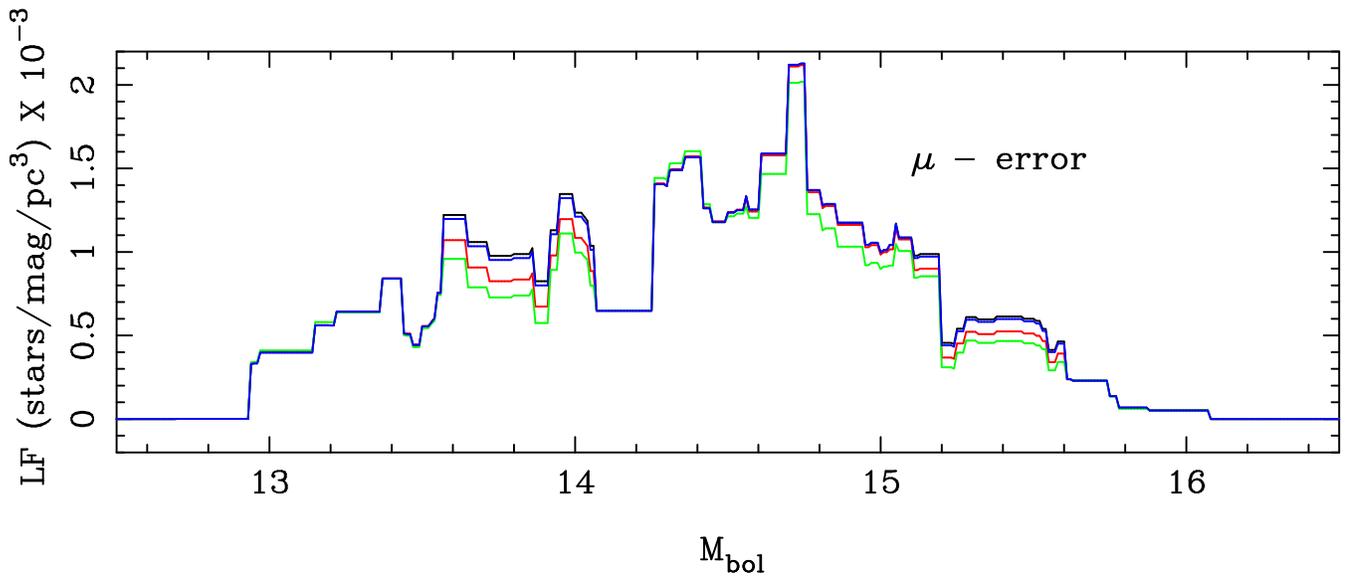
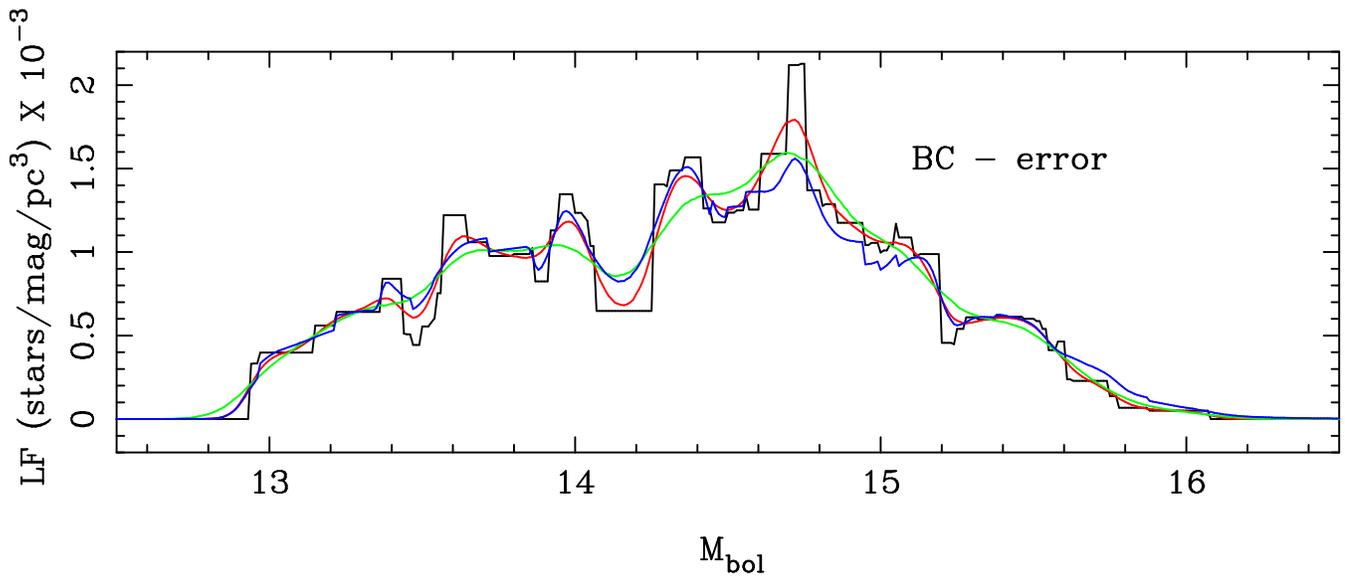
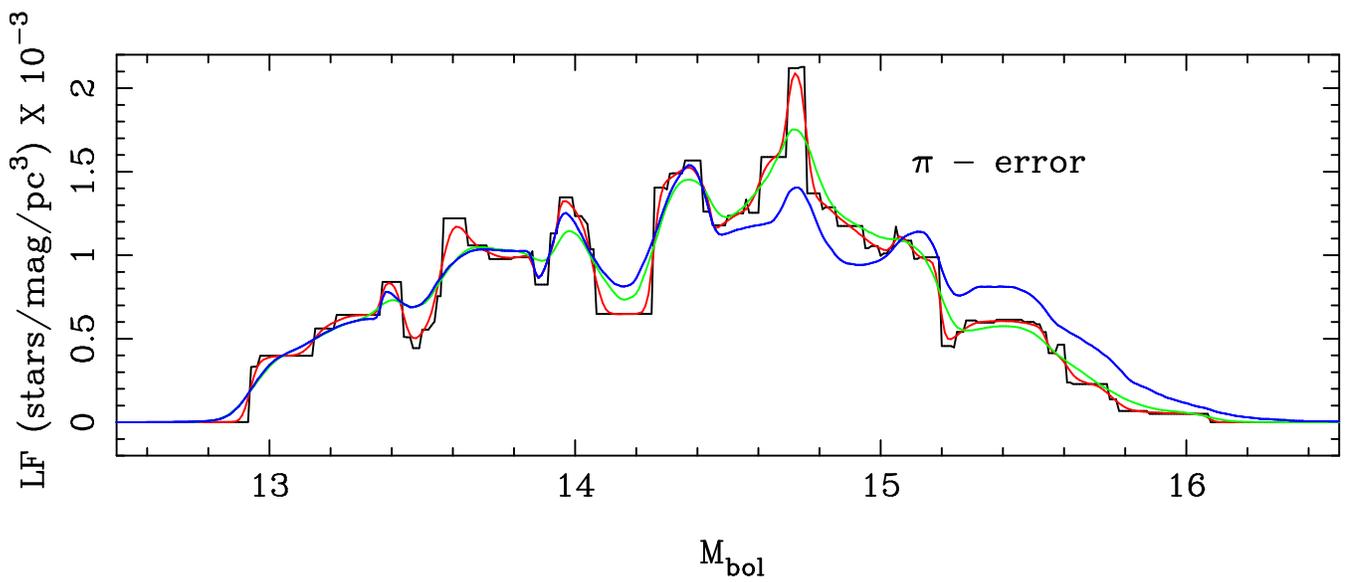
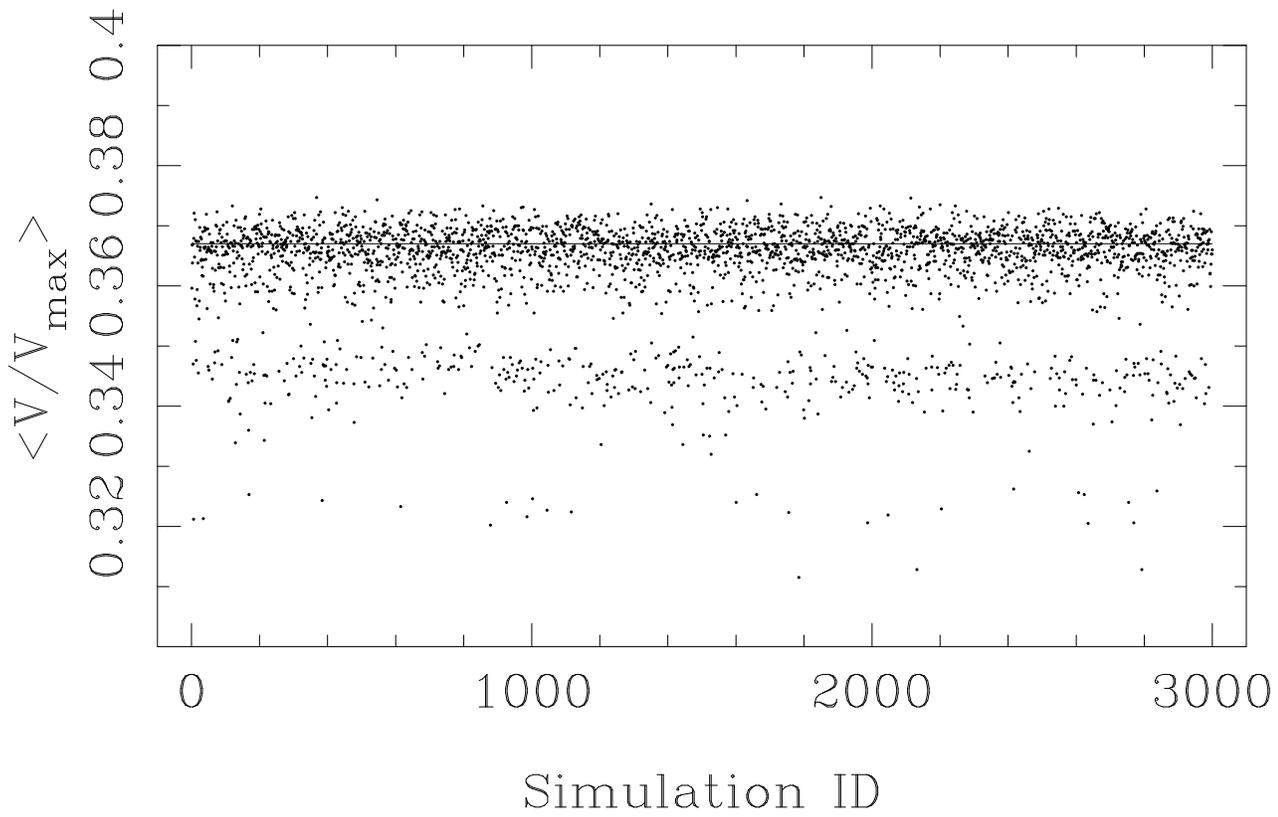
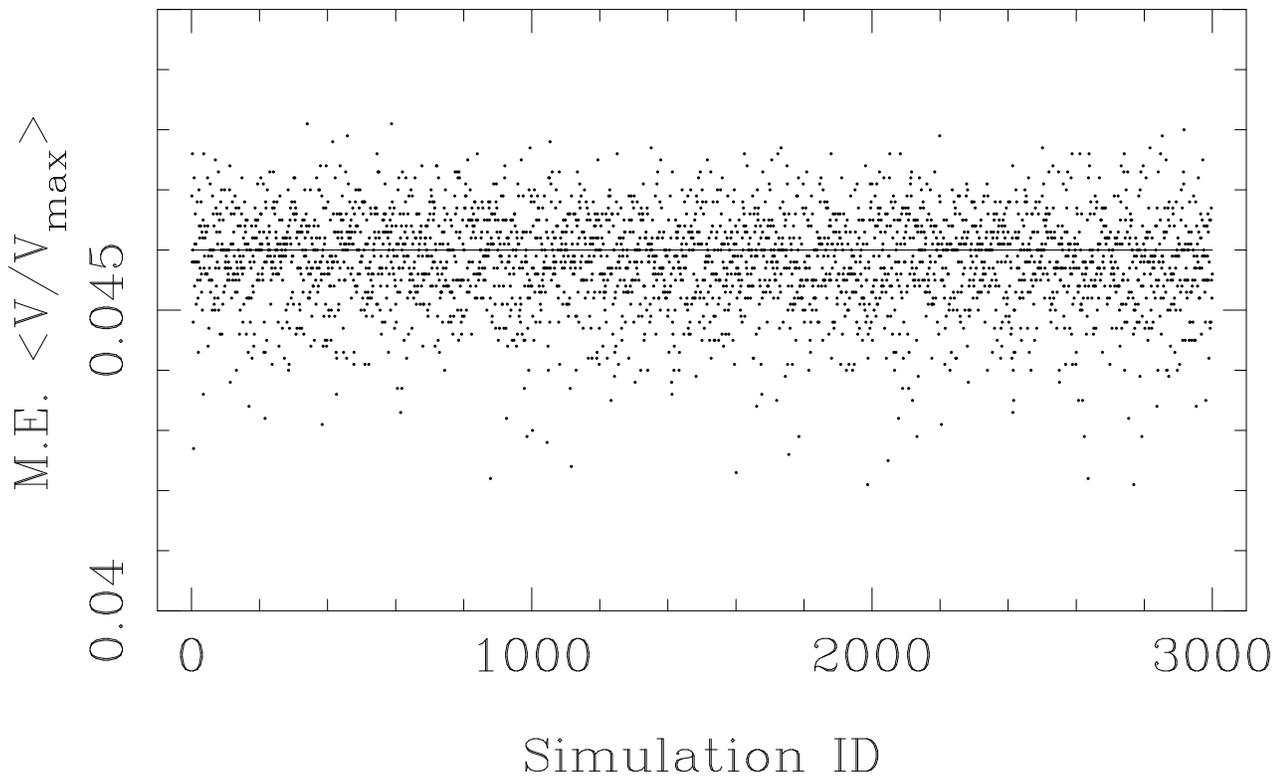
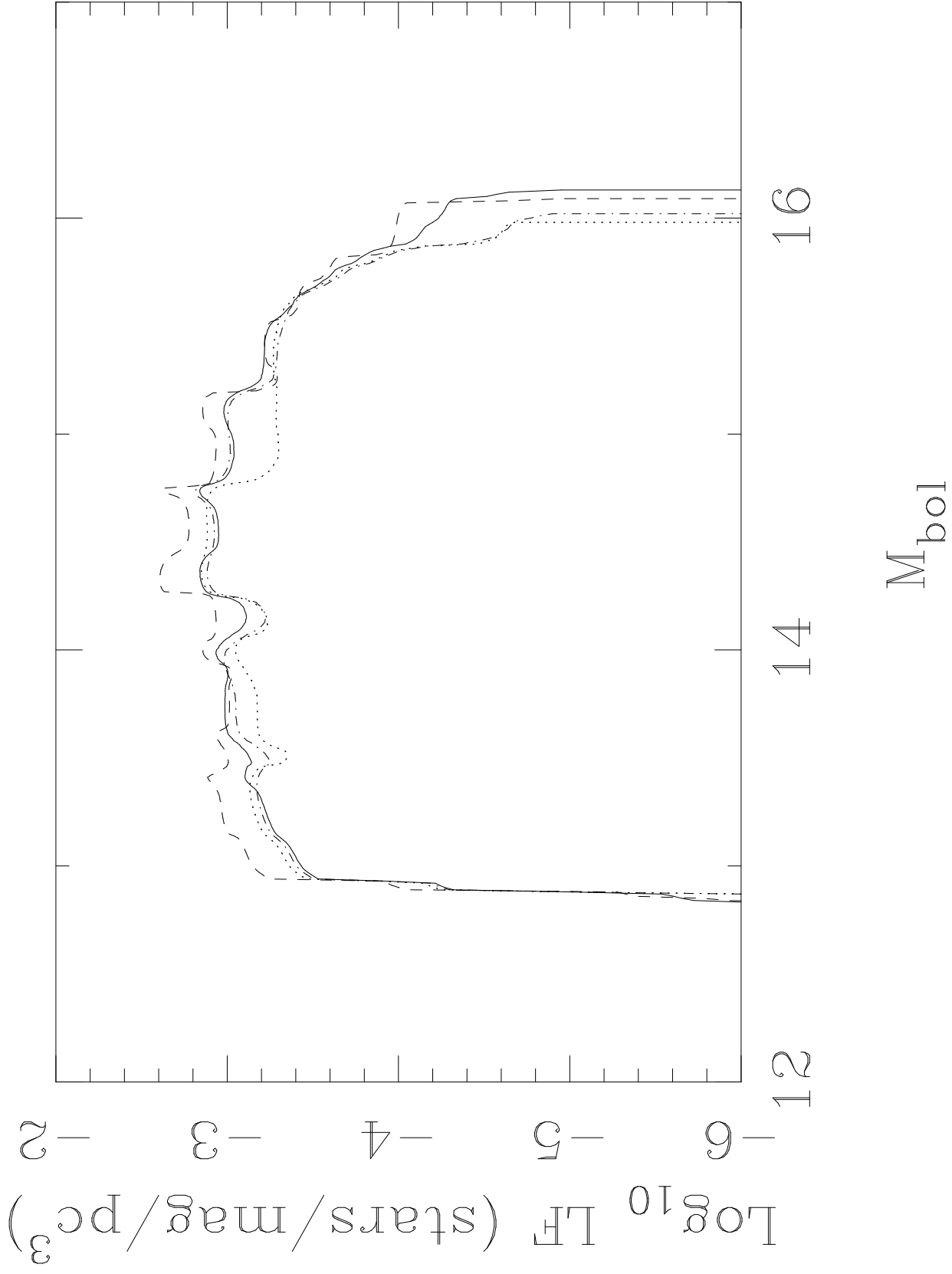
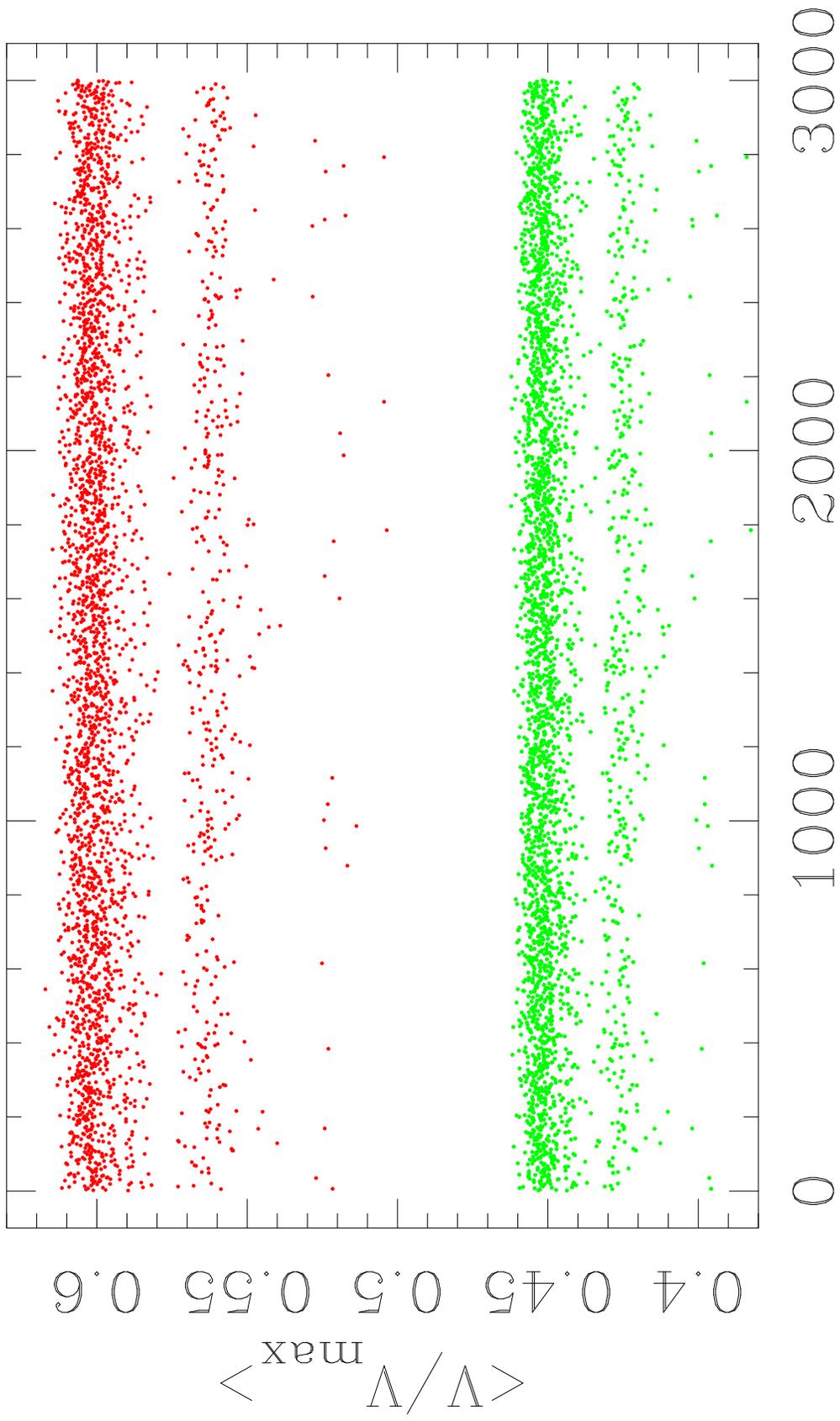
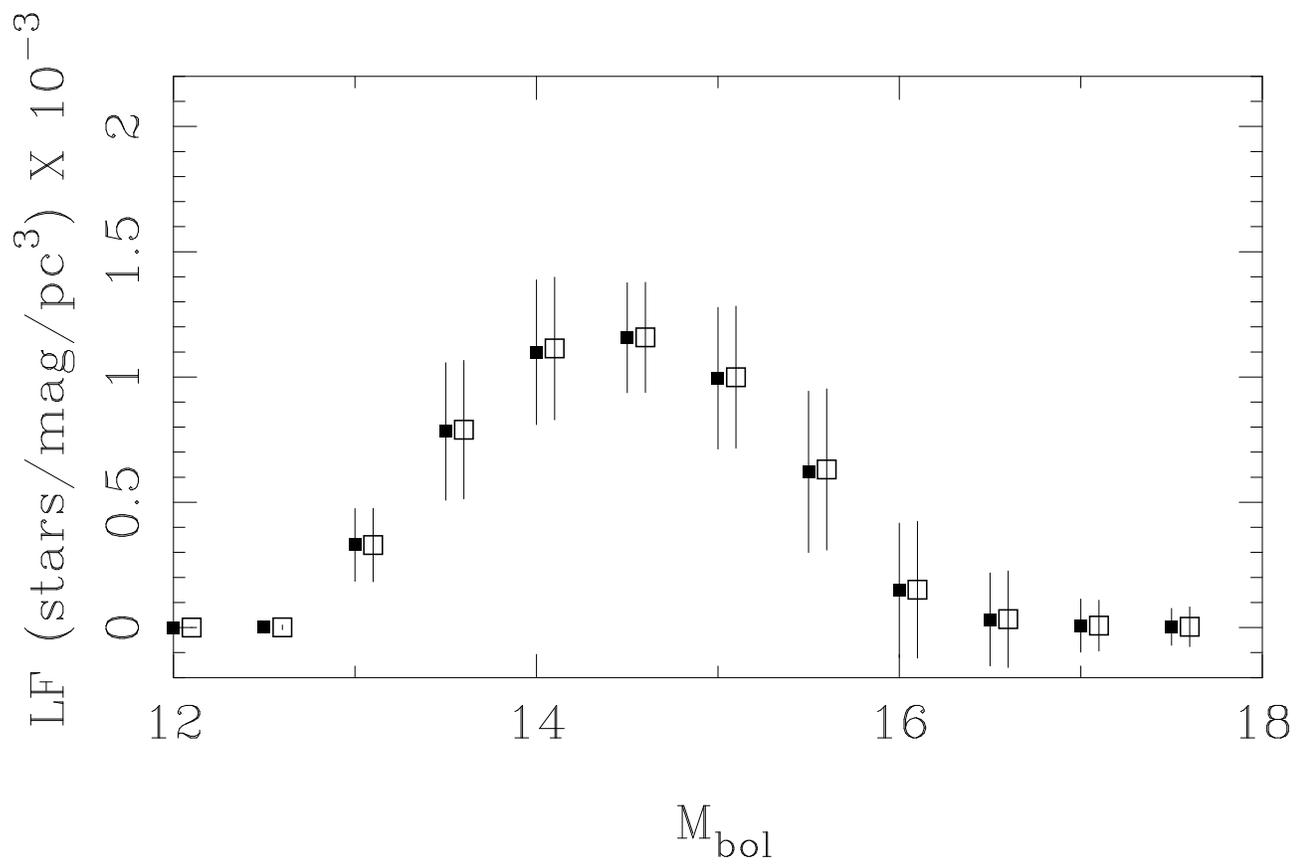
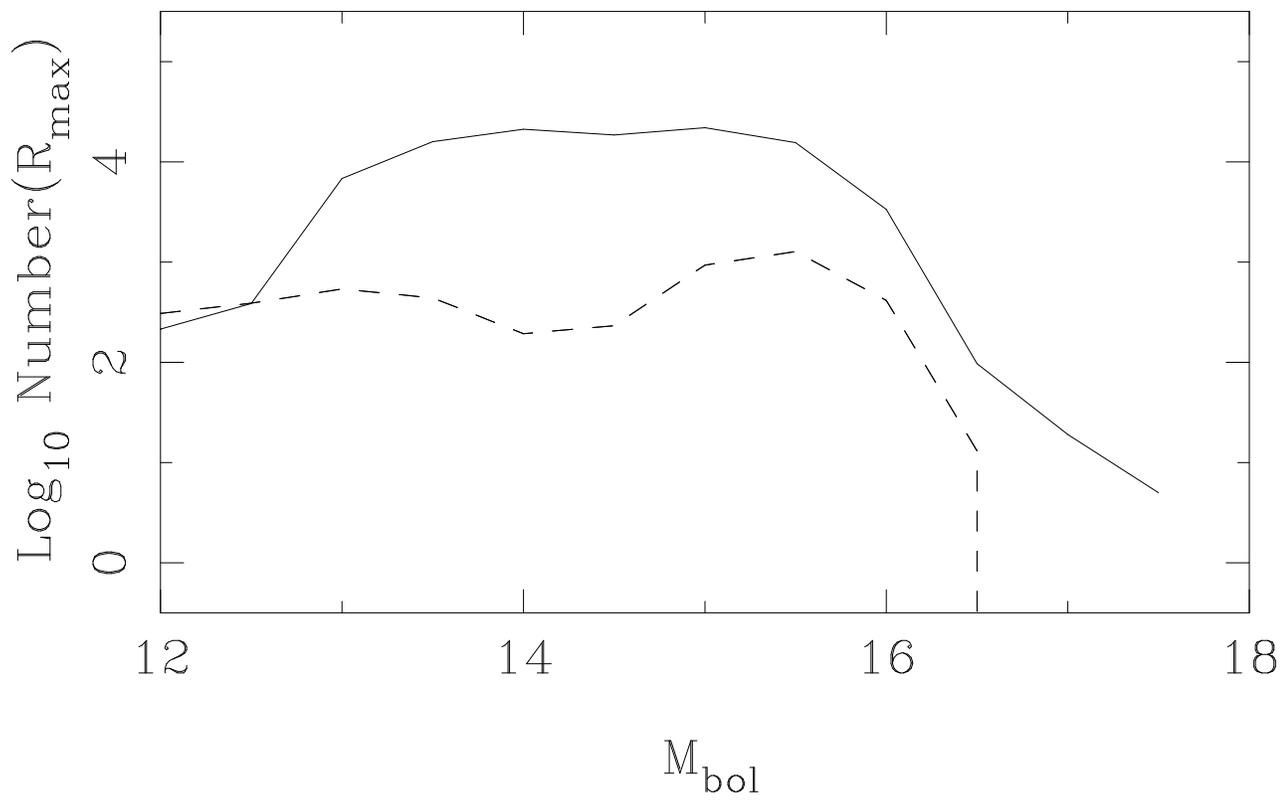
\end{document}